# Hyperbolic Graph Embeddings Reveal the Host–Pathogen Interactome

Xiaoqiong Xia[1-4] and Cesar de la Fuente-Nunez[1-4,*]

[1]Machine Biology Group, Departments of Psychiatry and Microbiology, Institute for Biomedical Informatics, Institute for Translational Medicine and Therapeutics, Perelman School of Medicine, University of Pennsylvania, Philadelphia, Pennsylvania, United States of America.

[2]Departments of Bioengineering and Chemical and Biomolecular Engineering, School of Engineering and Applied Science, University of Pennsylvania, Philadelphia, Pennsylvania, United States of America.

[3]Department of Chemistry, School of Arts and Sciences, University of Pennsylvania, Philadelphia, Pennsylvania, United States of America.

[4]Penn Institute for Computational Science, University of Pennsylvania, Philadelphia, Pennsylvania, United States of America.

*Correspondence:

Cesar de la Fuente-Nunez (cfuente@upenn.edu)

## Abstract

Infections depend on interactions between pathogen and host proteins, but comprehensively mapping these interactions is challenging and labor intensive. Many biological networks have hierarchical, scale-free structure, so we developed a deep learning framework, ApexPPI, that represents protein networks in hyperbolic Riemannian space to capture these features. Our model integrates multimodal biological data (protein sequences, gene perturbation experiments, and complementary interaction networks) to predict likely interactions between pathogen and host proteins through multi-task hyperbolic graph neural networks. Mapping protein features into hyperbolic space led to much higher accuracy than previous methods in predicting host–pathogen interactions. From tens of millions of possible protein pairs, our model identified thousands of high-confidence interactions, including many involving human G-protein-coupled receptors (GPCRs). We validated dozens of these predicted complexes using AlphaFold 3 structural modeling, supporting the accuracy of our predictions. This comprehensive map of host–pathogen protein interactions provides a resource for discovering new treatments and illustrates how advanced AI can unravel complex biological systems.

## Introduction

The physical interactions between host and pathogen proteins are fundamental to infection: pathogen proteins must bind host targets to hijack cellular machinery, evade immunity, and replicate[1]. Accordingly, characterizing the host–pathogen interactome is essential for understanding disease mechanisms and identifying intervention points. In particular, disrupting key HP-PPIs (e.g. blocking a virus protein from binding an immune receptor) has emerged as a promising therapeutic strategy)[2]. Among therapeutic modalities, short peptides are especially well suited to modulate PPIs at shallow, extended interfaces; recent advances in peptide design improved their stability and in vivo performance[3-5]. Despite this promise, known HP-PPI data remain sparse: public resources contain only thousands of experimentally validated interactions, and broad discovery via yeast two-hybrid screens or affinity capture is labor-intensive and often incomplete[6]. This experimental bottleneck motivates the development of computational methods to fill the gap of accurate, scalable predictions across diverse pathogen-host systems[7].

Recent advances in machine learning, especially deep learning, have indeed enabled rapid prediction of PPIs from diverse data[8]. Models such as DeepPPI[9], FGNN[10], and DSSGNN-PPI[11] have demonstrated significant predictive capability when trained on large-scale, multi-modal datasets from resources including HPIDB (experimentally validated HP-PPIs)[12], STRING (comprehensive PPI networks)[13], UniProt (protein sequences and annotations)[14], and LINCS (gene expression and perturbation profiles)[15]. These databases provide complementary perspectives: HPIDB supplies curated interaction gold standards, STRING contextualizes proteins within broader networks, UniProt enables sequence-based feature extraction, and LINCS captures functional responses to molecular perturbations.

Despite these advances, HP-PPI prediction faces unique challenges arising from the substantial heterogeneity of host and pathogen proteomes, the diversity of molecular interaction contexts[16], and the scarcity of experimentally validated training data[17]. Standard Euclidean graph embeddings struggle to efficiently represent the hierarchical, scale-free architecture characteristic of biological networks, where hub proteins interact with many partners while most proteins participate in few interactions[16]. Hyperbolic (Lorentzian) geometry offers a mathematically principled solution: by naturally accommodating hierarchical relationships and tree-like structures[17], hyperbolic embeddings require fewer dimensions than Euclidean alternatives while better preserving network topology. This geometric approach is particularly well-suited for sparse, heterogeneous HP-PPI data where capturing latent hierarchical organization is critical for accurate prediction. Furthermore, reconstructing complementary interaction networks, including intra-species pathogen-pathogen and human-human PPIs, sequence similarity networks, and gene perturbation networks, can enrich training data, reveal shared interaction motifs, and improve cross-context generalization. Multi-task learning across these related networks enables the model to leverage structural and functional patterns common across biological systems.

In this work, we present a hyperbolic graph neural network framework that integrates hyperbolic embeddings[18], multi-task learning[19], multi-modal data fusion[20], and ensemble modeling[21] to predict HP-PPIs with high accuracy and biological interpretability (**Figure 1**). Our approach learns Lorentzian-space embeddings for each protein, followed by a hyperbolic multi-layer perceptron that scores protein–protein pairs. We ensemble several variants of this model to maximize robustness. Crucially, predicted interactions are treated as high-priority candidates rather than definitive; we validate top-scoring pairs using AlphaFold3[22] to predict complex structures and Pyrosetta to compute binding energetics. The resulting pipeline yields a highly enriched set of candidate HP-PPIs (84 general cases plus 19 GPCR-related pairs) with credible structural support. Overall, our framework achieves state-of-the-art HP-PPI prediction (AUROC 0.905), discovers biologically plausible novel interactions (especially involving druggable host receptors), and provides a scalable AI-driven strategy for mapping infection-relevant networks.

## Results

**Sequence embeddings alone are insufficient for HP-PPI prediction.** We first encoded all proteins using the pretrained ESM-2 language model (esm2_t12_35M_UR50D) to obtain 480-dimensional embeddings. Analysis of protein length distributions revealed that 95% of host proteins and 98% of pathogen proteins contain fewer than 1,024 residues (**Fig. 2a,b**), validating our maximum sequence length threshold and minimizing truncation artifacts.

However, projecting these embeddings via UMAP revealed extensive overlap between host and pathogen proteins in sequence space (**Fig. 2c**), reflecting shared biochemical motifs but indicating that sequence features alone cannot distinguish species or predict interactions. Likewise, when we concatenated pairwise embeddings for protein pairs and projected them (**Fig. 2d**), interacting (positive) and non-interacting (negative) pairs showed no clear separation. These observations imply that raw sequence similarity is a weak signal for HP-PPI classification. In contrast, after running our hyperbolic GNN, the learned pairwise embeddings cleanly separated interacting from non-interacting pairs (**Fig. 2e**). This demonstrates that incorporating network topology into the embedding – by passing messages and aggregating neighbors on a hyperbolic manifold – transforms the raw sequence vectors into geometry-aware representations that effectively encode interaction likelihood.

**HP-PPI networks exhibit a scale-free topology consistent with hyperbolic embedding.** We next characterized the topology of the curated HP-PPI network (58,653 validated links among 13,100 host and 6,782 pathogen proteins). A force-directed layout (**Fig. 3a**) already suggests a classic hub-and-spoke architecture: a few large nodes (hubs with many interactions) occupy the center, connected to many small peripheral nodes (few interactions). Indeed, the degree distribution plotted on logarithmic scales is nearly linear (**Fig. 3b**), indicating a heavy-tailed, power-law form. Maximum likelihood fitting yielded an exponent $\alpha=2.43$ ($x_{\min}=16$), in the range typical of biological networks. Crucially, log-likelihood ratio testing strongly favored the power-law model over exponential alternatives (R=407.69, $p<0.0001$), confirming genuine scale-free organization rather than statistical artifact.

In addition, the average clustering coefficient of this network is nearly zero (**Table 1**), showing that high-degree hubs do not form tightly connected cliques – another hallmark of hierarchical, tree-like structure. Together, these results empirically justify using hyperbolic geometry: as noted in the literature, hyperbolic models naturally reproduce key features of real-world networks, including power-law degree distributions and hierarchical structure. We performed analogous analyses on the auxiliary networks used for multi-task training (**Table 1**). The pathogen sequence similarity network also showed a significant power-law tail ($\alpha=2.48$, $p<0.0001$), whereas the drug-perturbation and gene knockdown/overexpression networks had weaker scale-free signatures (high α, low R, high p) and moderate clustering, indicating some hierarchical structure but not strict scale-freeness. The host sequence similarity graph was essentially complete (high average clustering) at our threshold, explaining the atypically large fitted α ("saturated"). Overall, most networks incorporated have at least partial hierarchical organization, supporting the choice of hyperbolic embedding as a geometric prior in learning.

Network visualization using force-directed layout (**Fig. 3a**) illustrates the characteristic hub-and-spoke architecture, with central high-degree nodes surrounded by peripheral low-degree proteins. This topological analysis provides strong empirical justification for hyperbolic embedding as the geometric prior most aligned with HP-PPI network structure.

**Ensemble modeling optimizes predictive accuracy.** To maximize prediction accuracy while avoiding overfitting, we systematically evaluated ensemble model performance by incrementally aggregating predictions from individual architectures with varying hyperparameters. Both AUROC and AUPR increased rapidly during initial ensemble aggregation, with AUROC peaking at 0.905 and AUPR reaching 0.703 (**Fig. 4**). Performance gains were most pronounced for the first six models, suggesting early members contribute maximally diverse and complementary information. Beyond six models, further additions yielded diminishing returns followed by gradual performance decline, likely due to redundancy or noise amplification. Based on this analysis, we selected the top six architectures (**Table S2**) as our final ensemble, balancing predictive power with computational efficiency. These models varied in learning rate (0.0001–0.001), hidden feature dimensionality (128–256), and multi-task loss weights, ensuring architectural diversity that enhances robustness and generalization.

**Multi-stage structural validation filters high-condidence predictions.** To identify biologically relevant HP-PPIs from the vast combinatorial space, we implemented a stringent multi-stage filtering pipeline (**Fig. 5a**). From 88,790,128 possible host-pathogen protein pairs, our ensemble model identified 6,806 high-confidence candidates with prediction scores >0.99, representing the top 0.008% of all evaluated pairs (**Fig. 5b**). These candidates exhibited strongly right-skewed score distribution (median = 0.996), with pronounced enrichment near the maximum score, indicating robust model certainty. Recognizing the therapeutic importance of G protein-coupled receptors (GPCRs), we additionally evaluated 1,900 GPCR-pathogen interactions that achieved scores >0.7, applying a lower threshold to capture potentially relevant yet moderately confident predictions in this druggable target class. To assess structural plausibility, all high-scoring candidates underwent AlphaFold3 structural modeling, which predicts binding interface geometry and confidence. Using an interface predicted TM-score (ipTM) cutoff of >0.6, indicating reliable interface positioning, we identified 84 general HP-PPIs and 19 GPCR-pathogen interactions with credible structural support. The ipTM distribution (**Fig. 5c**) for these validated complexes showed values concentrated between 0.60 and 0.75, confirming well-formed binding interfaces. This structural validation step effectively filtered the initial 8,706 computational predictions (6,806 + 1,900) to 103 structurally compatible complexes, representing a critical quality control that eliminates geometrically implausible interactions. To further evaluate binding stability, all 103 structurally validated complexes were refined using PyRosetta's FastRelax protocol with the REF2015 energy function. Interface binding energies were computed to assess thermodynamic favorability. The Rosetta energy distribution (**Fig. 5d**) revealed that all validated complexes exhibited favorable (negative) binding energies, spanning from -4,000 to -500, with the majority showing strongly negative values indicative of stable interactions. This energetic refinement provides additional orthogonal evidence supporting the biological relevance of predicted interactions.

Collectively, the convergence of three independent metrics, including high graph-based prediction scores (**Fig. 5b**), confident interface geometry from AlphaFold3 (**Fig. 5c**), and favorable binding energetics from Rosetta (**Fig. 5d**), establishes a high-confidence set of 103 HP-PPIs (84 general + 19 GPCR-pathogen) warranting experimental validation. These multi-layered validation criteria ensure that prioritized candidates exhibit not only compatible network topology but also physically realizable interfaces with thermodynamically stable binding, substantially enriching for genuine protein-protein interactions.

## Methods

### Data collection and preprocessing

**Host-pathogen interaction data.** We obtained experimentally validated HP-PPI data from HPIDB 3.0, initially comprising 69,787 interaction pairs. After removing duplicates, the curated dataset contained 58,653 unique HP-PPIs involving 13,100 host proteins and 6,782 pathogen proteins. Each protein entry includes UniProt identifiers, taxonomic classifications, amino acid sequences, and network node indices (**Table S1**).

**Human protein interactions.** To augment training data with intra-species interactions, we extracted human PPIs from STRING v11.5 using a confidence score threshold of 400 (medium confidence). From 1,858,944 human PPIs, we constructed a subnetwork containing only proteins present in the HP-PPI dataset, yielding 7,140 shared proteins and 611,478 interactions. Protein identifiers were mapped to Ensembl gene IDs for consistency. Both HP-PPI and human PPI networks were converted to zero-indexed node pair format for computational efficiency.

**Sequence similarity networks.** We computed pairwise sequence similarity using the Smith-Waterman local alignment algorithm with the BLOSUM50 substitution matrix. Protein pairs achieving similarity scores >0.7 were connected in sequence similarity networks, constructed separately for host (PPI_H_sim) and pathogen (PPI_P_sim) proteins. This threshold balances sensitivity (capturing homologous relationships) with specificity (avoiding spurious connections).

**Gene perturbation networks.** We downloaded gene expression perturbation profiles from the LINCS L1000 dataset, encompassing responses to small molecule drugs, gene overexpression, and gene knockdown. We transposed perturbation matrices such that each gene is represented by its response signature across perturbations. For each perturbation type (drug, overexpression, knockdown), we computed pairwise Pearson correlation coefficients and applied a threshold of 0.7 to define edges. Gene identifiers were mapped to host proteins in the HP-PPI dataset, generating three perturbation-based networks: PPI_DR (drug response), PPI_OE (overexpression), and PPI_KD (knockdown).

**Protein sequence embeddings**
We generated initial protein feature vectors using the ESM-2 pretrained model (esm2_t12_35M_UR50D), a protein language model pretrained on 65 million sequences. ESM-2 encodes each amino acid as a 480-dimensional vector capturing evolutionary and structural information. For proteins exceeding 1,024 residues (5% of host proteins, 2% of pathogen proteins), we applied truncation. Mean pooling across residue embeddings yielded a single 480-dimensional vector per protein, producing a 19,882 × 480 feature matrix used as initial node features in graph neural networks.

**Network topology analysis**
To characterize network structure, we computed degree distributions and tested for scale-free topology using maximum likelihood power-law fitting. For each network, we estimated the scaling exponent α and minimum degree threshold $x_{min}$ above which power-law behavior emerges. Statistical support was assessed via log-likelihood ratio tests comparing power-law against exponential distributions. Networks with R > 0 and $p < 0.05$ were classified as scale-free. Clustering coefficients were computed to assess local connectivity patterns.

**Hyperbolic graph neural networks**
**Lorentz model.** We embedded PPI networks in the Lorentz model of hyperbolic space, which represents points on a hyperboloid satisfying $<x,x>_L = -1$ in Minkowski spacetime. This geometry naturally accommodates hierarchical structures by enabling exponential volume growth with radius, contrasting with polynomial growth in Euclidean space.

**Hyperbolic graph convolutions.** At each layer, we transform node features to the tangent space at the origin, perform neighborhood aggregation in the tangent space, and project back to the hyperbolic manifold. Formally, for node $v$ at layer $k$:
1. **Map to tangent space:** Transform neighbor embeddings from hyperbolic manifold to Euclidean tangent space via logarithmic map $h^T = log_o(h^H)$, where $h^T$ is the hidden vector in tangent space, and $h^H$ is the hidden vector in hyperbolic space.
2. **Aggregate neighbors:** Compute mean of neighbor features in tangent space $h_N(v) = mean(\{log_o(h_u^H): u \in N(v)\})$.
3. **Aggregate neighbors:** Compute mean of neighbor features in tangent space:
$$h_v^{k+1} = exp_o(W^k \cdot h_v^k + W_N^k \cdot h_N(v) + b^k)$$

Where $log_o$ and $exp_o$ denote exponential and logarithmic maps at the origin, $W^k$ are learnable weight matrices, and $b^k$ are bias terms. We used curvature c = -1 and two-layer architectures ($k = 2$).

**Link prediction.** For each protein pair $(u, v)$, we concatenated their hyperbolic embeddings and passed the result through a hyperbolic multi-layer perceptron:
$$z_{uv} = concat(h_u, h_v)$$
$$s_{uv} = MLP^H(z_{uv})$$
where $MLP^H$ consists of hyperbolic linear transformations alternating with LeakyReLU activation (negative slope = 0.5). Binary cross-entropy loss was used for training: $\mathcal{L} = -[y\log(\sigma(s_{uv})) + (1-y)\log(1-\sigma(s_{uv}))]$, where $y \in [0,1]$ indicates ground truth interaction status and $\sigma$ denotes the sigmoid function.

**Multi-task learning**

To leverage complementary information from auxiliary networks, we implemented multi-task learning with shared hyperbolic encoders. For each network type m ∈ {HP-PPI, STRING, sequence similarity, perturbations}, we computed task-specific embeddings and combined them additively:
$$h_{final} = \sum_m h_m$$
where each $h_m$ is learned via the hyperbolic GNN applied to network m. Task-specific losses were weighted and summed: $\mathcal{L}_{total} = \sum_m w_m \mathcal{L}_m$, with weights $\{w_m\}$ determined via hyperparameter search. This approach enables the model to extract shared interaction principles while respecting task-specific patterns.

**Training and validation**

We performed 10-fold cross-validation to assess generalization. For each fold, we split data into training (80%), validation (10%), and test (10%) sets. Negative samples were generated by randomly pairing non-interacting proteins at ratios of 1:1, 1:3, 1:5, 1:10, 1:20, and 1:30 (positive:negative). The 1:10 ratio provided optimal balance between class imbalance and computational cost. Models were trained using Adam optimizer with learning rates in {0.0001, 0.001}, hidden dimensions in {128, 256}, and task-specific loss weights in {0.1, 1.0}. Early stopping based on validation AUROC prevented overfitting. The best six models (Table S2) were ensembled by averaging predicted probabilities.

**Structural validation with AlphaFold3**

High-scoring predictions (score > 0.99 for HP-PPIs, score > 0.7 for GPCR interactions) were subjected to AlphaFold3 modeling to assess structural plausibility. We computed interface predicted TM-scores (ipTM), which quantify confidence in relative chain positioning. Complexes with ipTM > 0.6 indicate moderate-to-high confidence in interface geometry and were retained for further analysis.

**Rosetta energy calculations**

AlphaFold3-predicted complexes were refined using PyRosetta FastRelax protocol with the REF2015 energy function. Interface binding energies were computed as $\Delta G = E_{complex} -$ ($E_{protein1} + E_{protein2}$), where negative values indicate favorable interactions. Structures exhibiting both high ipTM and negative Rosetta energies represent high-confidence physical interactions warranting experimental validation.

**Statistical analysis**

Performance metrics (AUROC, AUPR) were computed using scikit-learn. Significance of power-law fits was assessed via likelihood ratio tests with bootstrapped *p*-values (n=1,000 replicates). All analyses were performed in Python 3.8 using PyTorch 1.12, DGL 0.9, and NetworkX 2.8.

## Discussion

We have presented a geometric deep learning framework that predicts host–pathogen protein interactions by embedding multi-modal biological networks in hyperbolic space. Our approach substantially outperforms Euclidean baselines, achieving AUROC=0.905 and AUPR=0.703 on held-out HP-PPI test data. The success of hyperbolic embeddings highlights a key insight: biological interaction networks are inherently hierarchical and scale-free, a structure that Euclidean methods struggle to capture efficiently. Our analysis confirmed that the HP-PPI network has a heavy-tailed degree distribution ($\alpha \approx 2.43$) and minimal clustering, consistent with a tree-like architecture. Hyperbolic space, with its exponential volume growth, is a natural geometry for such networks and yields more compact representations than Euclidean space. Notably, some auxiliary networks (e.g. drug perturbations) were not strictly scale-free, yet hyperbolic embedding still provided gains. This suggests that hyperbolic GNNs can represent a variety of complex topologies, from modular co-expression graphs to nearly complete similarity networks, by simultaneously encoding hierarchical and community structure. In practice, the shared hyperbolic encoder successfully distilled useful patterns from all tasks.

Our multi-task strategy contributed to the model's generalization: learning from human PPIs, sequence homology, and gene expression correlations allowed the network to recognize underlying constraints on protein compatibility. For example, proteins that are highly similar in sequence or that respond similarly to perturbations often share binding propensities, even across species. By jointly training on these signals, the model learned latent features that transcend any single context.

A key strength of our pipeline is the combination of predictive modeling with structural validation. While the GNN efficiently narrows an enormous search space (>88 million pairs) to a manageable shortlist, the AlphaFold3 and Rosetta steps provide mechanistic verification. An AlphaFold3 ipTM >0.6 indicates the predicted chains form a coherent interface. We found that 84 host–pathogen pairs and 19 GPCR–pathogen pairs passed this structural filter, out of 8,706 candidates (approximately 1.2%). All of these also had favorable (negative) Rosetta ΔG, further suggesting stable binding. The convergence of high ensemble score, good interface geometry, and low ΔG makes these 103 interactions especially compelling. Importantly, they include several biologically interesting hypotheses. For instance, we identified specific GPCRs that may bind pathogen proteins; since GPCRs are highly druggable, these cases are prime candidates for follow-up.

This work has limitations. First, predictions are general and do not account for context such as tissue expression, subcellular localization, or infection stage. Future models could incorporate such layers (for example, by conditioning on cell-type-specific expression networks) to refine context-specific PPI predictions. Second, the hyperbolic GNN captures global topology but may miss fine-grained structural details (e.g. small binding motifs). Hybrid models that combine graph topology with sequence-based attention or structural priors might improve this. Third, all predictions require experimental validation. Our pipeline uses AlphaFold3 and Rosetta as proxies for physical binding, but true biochemical assays (co-immunoprecipitation, affinity measurements) are needed to confirm these interactions. We are currently working with collaborators to test the highest-scoring complexes in relevant host–pathogen systems. Finally, while hyperbolic embeddings provide geometric intuition (e.g. distance from the origin can reflect node "hubness"), interpreting model decisions remains an open challenge. Tools like integrated gradients or SHAP could be explored to identify which protein features most drive a given predicted interaction.

In conclusion, our hyperbolic graph learning framework offers a powerful, geometry-aware approach to mapping host–pathogen interactomes. By respecting the hierarchical topology of biological networks and integrating diverse data types, we achieve high predictive accuracy and identify novel interaction candidates that are structurally plausible and energetically favorable. This AI-driven pipeline can guide targeted experiments, helping to fill gaps in the HP-PPI map. More broadly, as biological data continue to grow in scale and complexity, geometric deep learning methods like the one presented here will be invaluable for revealing the hidden organization of life's molecular networks and for accelerating the discovery of new therapeutic interventions.

## Acknowledgements

Cesar de la Fuente-Nunez holds a Presidential Professorship at the University of Pennsylvania. Research reported in this publication was supported by the NIH R35GM138201 and DTRA HDTRA1-21-1-0014. We thank de la Fuente Lab members for insightful discussions.

## Conflict of interest

Cesar de la Fuente-Nunez is a co-founder of, and scientific advisor, to Peptaris, Inc., provides consulting services to Invaio Sciences, and is a member of the Scientific Advisory Boards of Nowture S.L., Peptidus, European Biotech Venture Builder, the Peptide Drug Hunting Consortium (PDHC), ePhective Therapeutics, Inc., and Phare Bio.

# Figures

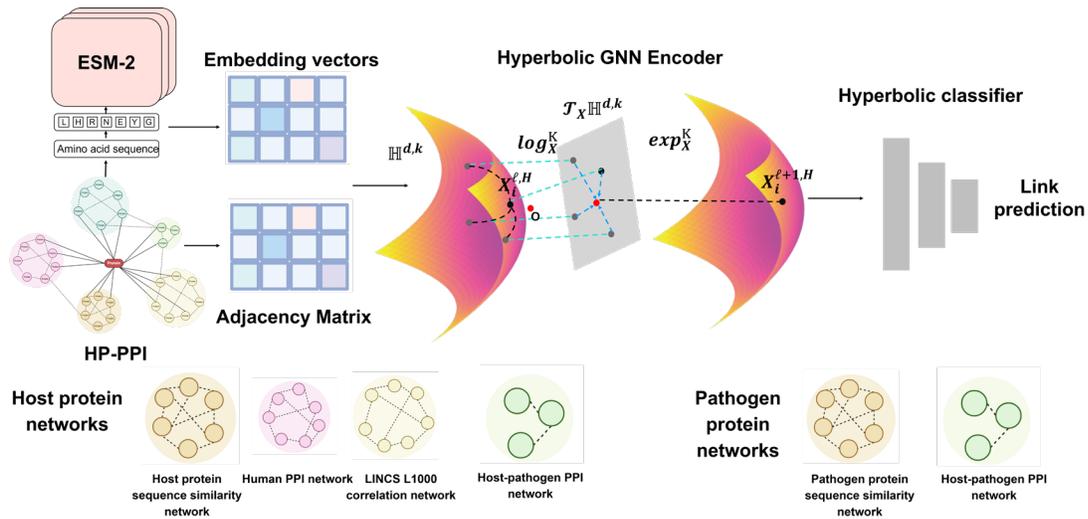

**Figure 1. Hyperbolic graph learning framework for HP-PPI prediction.** Host and pathogen protein sequences are first encoded by the ESM-2 language model into 480-dimensional embeddings that serve as initial node features. These serve as initial node embeddings for a hyperbolic graph neural network. Each graph convolutional layer maps node features to the tangent space of a Lorentzian hyperboloid, aggregates neighbor information by averaging tangent vectors, and maps the result back to the hyperbolic manifold. The updated hyperbolic embeddings capture both sequence-derived features and network topology. For link prediction, embeddings of protein pairs are concatenated and passed through a hyperbolic multi-layer perceptron, which outputs an interaction likelihood score via a sigmoid function.

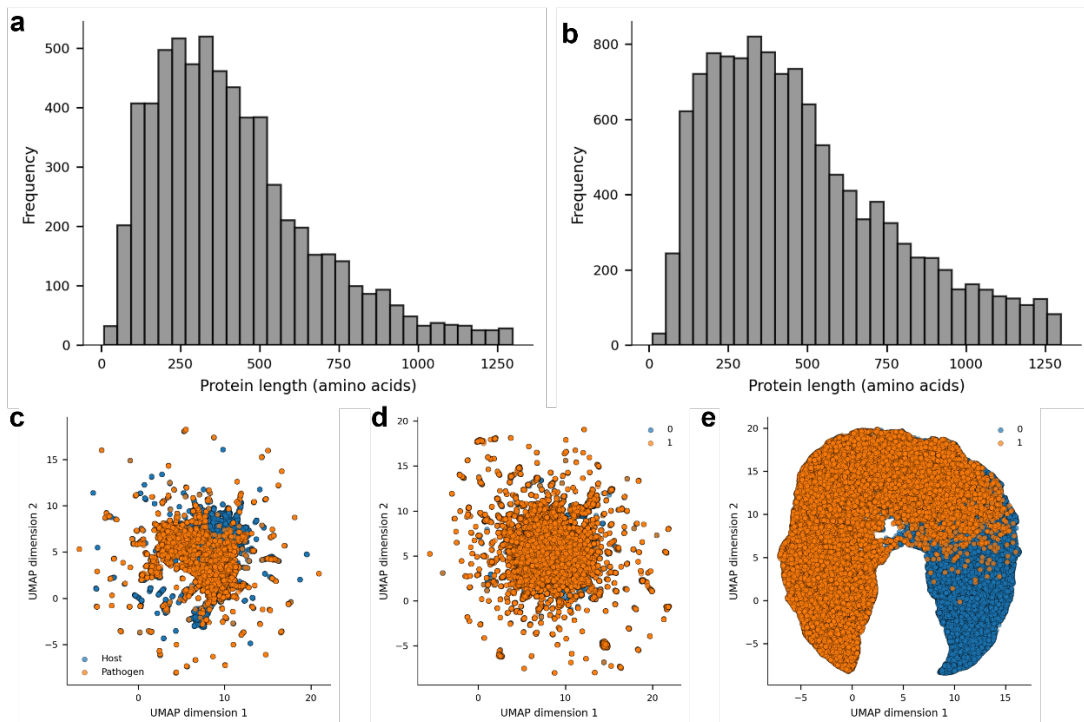

**Figure 2. Protein sequence embeddings capture biochemical features but require network integration for interaction prediction. (a, b)** Distributions of protein lengths in the HP-PPI dataset: **(a)** host proteins (n=13,100), **(b)** pathogen proteins (n=6,782). Over 95% of proteins in each set have ⩽1,024 amino acids, validating our truncation threshold for ESM-2 input. **(c)** UMAP projection of the 480-dimensional ESM-2 embeddings for host (blue) and pathogen (orange) proteins. The substantial overlap indicates that sequence-based representations alone do not distinguish host vs. pathogen origin, reflecting shared structural motifs and conservation. **(d)** UMAP of concatenated sequence embeddings for protein pairs labeled by interaction status: interacting (positive, orange) vs. non-interacting (negative, blue). The two classes remain intermixed, confirming that Euclidean sequence features alone do not separate true HP-PPIs. **(e)** UMAP of the hyperbolic graph embeddings for the same protein pairs after HGNN message passing. Here, interacting pairs cluster separately from non-interacting pairs, illustrating that incorporating network topology in hyperbolic space creates geometry-aware representations that discriminate interactions.

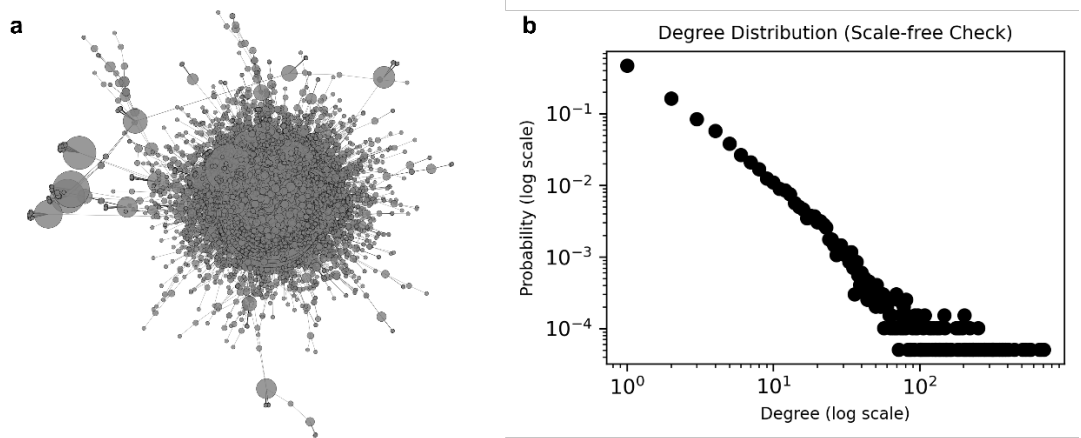

**Figure 3. HP-PPI networks exhibit scale-free topology supporting hyperbolic representation.**
**(a)** Force-directed layout of the HP-PPI network. Node size is proportional to degree (number of interactions). Large central nodes (hubs) connect to many other proteins, forming a "hub-and-spoke" architecture characteristic of hierarchical networks. Peripheral nodes are small (low degree). Edge opacity is reduced to prevent clutter. **(b)** Degree distribution plotted on log-log axes. The near-linear trend for higher degrees indicates a power-law (heavy-tailed) distribution. Maximum-likelihood fitting yields exponent $\alpha=2.43$ (for degrees $\geq 16$), and a likelihood-ratio test strongly favors this power-law model over an exponential model ($R=407.69$, $p<0.0001$). These results confirm the network's scale-free, hierarchical nature**.**

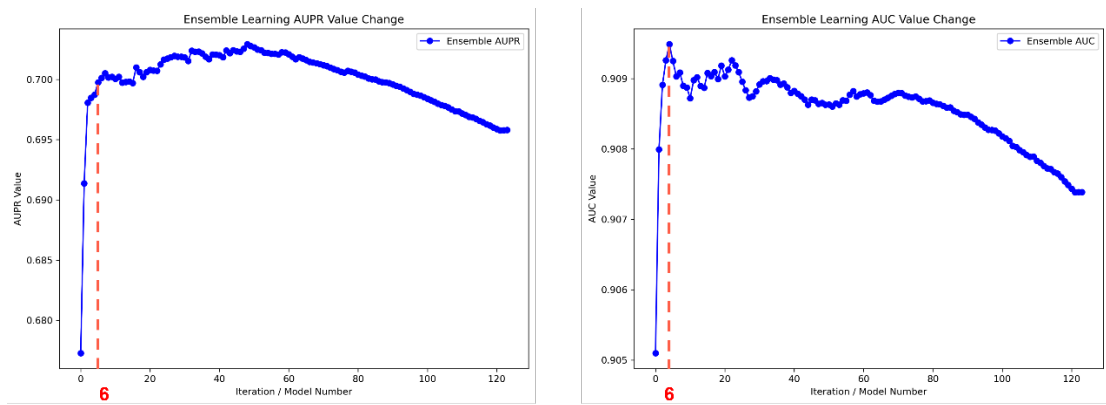

**Figure 4. Ensemble aggregation optimizes predictive performance. (a, b)** Area under the precision-recall curve (AUPR, **panel a**) and under the receiver operating characteristic (AUROC, panel **b)** as functions of ensemble size. Models were added one at a time in descending order of individual validation performance. Performance improves markedly as the first several models are aggregated, reaching a maximum at six models (AUROC=0.905, AUPR=0.703). Adding more models yields diminishing returns and eventually degrades accuracy, reflecting redundancy. This pattern indicates that a small ensemble of diverse models suffices to capture the majority of predictive power

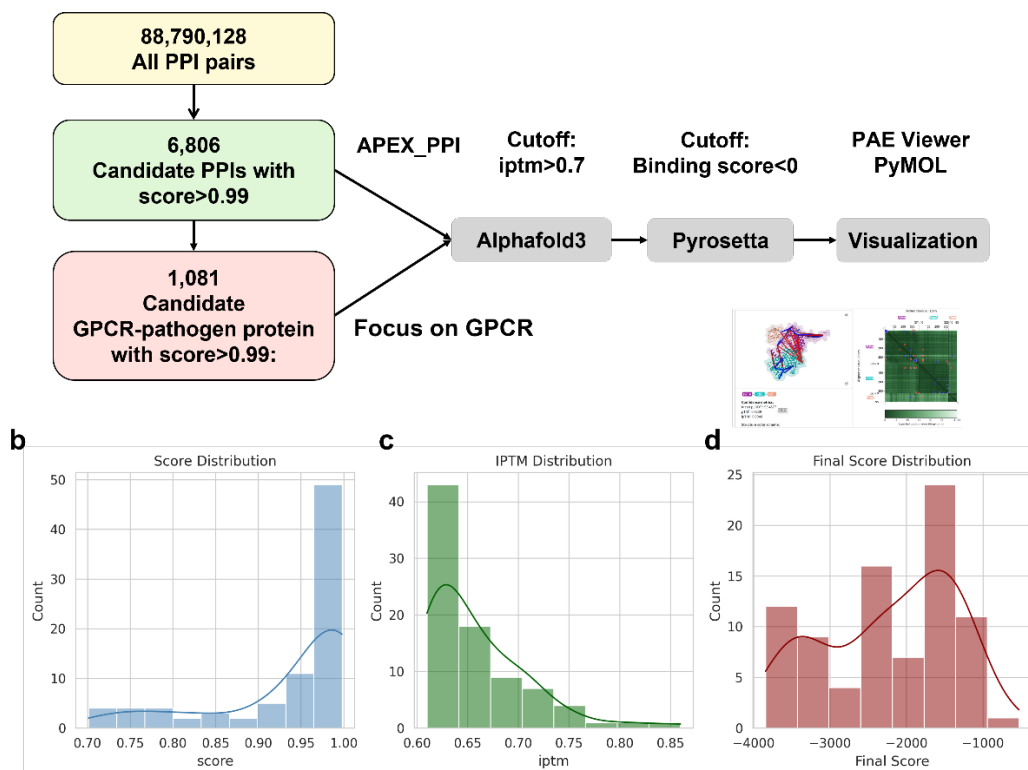

**Figure 5. Multi-stage validation pipeline identifies high-confidence HP-PPIs. (a)** Workflow diagram. All 88.79 million possible host–pathogen protein pairs were first scored by the HGNN

ensemble. Pairs with score >0.99 (6,806 host–pathogen and 1,900 GPCR–pathogen candidates) were selected as high-confidence predictions. Each candidate pair was then modeled with AlphaFold3 to predict the complex structure; complexes with interface TM-score (ipTM) > 0.6 were retained. Finally, retained complexes were energy-minimized with PyRosetta FastRelax and binding energies (ΔG) were computed; only interactions with favorable (negative) ΔG were kept. **(b)** Distribution of ensemble prediction scores for the 6,806 top-scoring HP-PPI candidates. The distribution is highly right-skewed (median = 0.996) with a sharp peak near 1.0, indicating strong model confidence in these predictions. **(c)** Distribution of AlphaFold3 interface-predicted TM-scores (ipTM) for the structurally modeled complexes. All shown complexes exceed the 0.6 cutoff, confirming well-formed interfaces. **(d)** Distribution of Rosetta binding energies (ΔG) after structural refinement, plotted on a log scale. All 103 validated complexes have negative ΔG (range ≈ –4,000 to –500 kcal/mol), indicating thermodynamically favorable interactions.

# Tables

**Table 1. Topological characterization of biological networks used in multi-task training.** Each network's degree distribution was tested for power-law behavior using maximum likelihood fitting. For each network we report the fitted exponent α, the log-likelihood ratio statistic R (positive values favor power-law), the *p*-value (statistical support for scale-free behavior), and the average clustering coefficient C. * denotes $p<0.0001$. The host–pathogen PPI network ("HP-PPI") has α=2.43 and R=407.69 ($p<0.0001$), confirming a scale-free, hierarchical topology. The pathogen sequence similarity network is similarly scale-free (α=2.48, $p<0.0001$) but with higher clustering (C=0.51), reflecting modular substructure. The drug perturbation (PPI_DR) and knockdown (PPI_KD) networks do not significantly fit a power law and have higher clustering, indicating more uniform (clustered) structure. The overexpression network (PPI_OE) shows weak hierarchy ($p\approx0.16$). The host similarity graph is nearly complete at our threshold (very high α and C≈0.47). In summary, most networks exhibit at least partial scale-free/hierarchical features (supporting hyperbolic embedding), except for the saturated host similarity graph.

| Network | Power law $\alpha$ | R statistic | p-value | Avg. clustering | Interpretation |
| --- | --- | --- | --- | --- | --- |
| **HP-PPI** | 2.43 | 407.69 | <0.0001* | <0.1 | Scale-free, hierarchical |
| **Pathogen similarity** | 2.48 | 373.87 | <0.0001* | 0.51 | Scale-free with modularity |
| **PPI_DR** | 22.89 | -0.03 | 0.857 | 0.59 | Non-scale-free, clustered |
| **PPI_KD** | 4.15 | -2.60 | 0.093 | 0.60 | Weakly hierarchical |
| **PPI_OE** | 4.50 | -1.49 | 0.160 | 0.54 | Weakly hierarchical |
| **Host similarity** | 789.46 | - | - | 0.47 | Saturated distribution |

*$p < 0.0001$ indicates strong statistical support for power-law over exponential model.